\documentclass{ws-ijmpa}

\begin{document}

\markboth{G.S. Huang}
{Observation of New Hadronic $\psi(2S)$ Decays at CLEO}

%
\catchline{}{}{}{}{}
%

\title{Observation of New Hadronic $\psi(2S)$ Decays at CLEO}

\author{G.S. Huang \\ Representing the CLEO Collaboration}

\address{Physics Department, Purdue University, West Lafayette, IN 47907, USA}

\maketitle


\begin{abstract}
   Using 5.46 pb$^{-1}$ of $e^+e^-$ annihilation data
   collected at the $\psi(2S)$ with the CLEO detector
   we have observed a variety of new hadronic decay
   modes of the $\psi(2S)$. A comprehensive set of
   branching ratios and upper limits is presented.

\end{abstract}

\section{Introduction}

In perturbative QCD the states $J/\psi$ and $\psi(2S)$ are 
non-relativistic bound
states of a charm and an anti-charm quark. The decays of these states are expected to be
dominated by the annihilation of the constituent $c\bar{c}$ into three
gluons. The partial width for the decays into an exclusive hadronic state, $h$,
is expected to be proportional to the square of the 
$c\bar{c}$ wave function overlap at the origin, which is well
determined from the leptonic width~\cite{PDG}. Since the strong coupling 
constant, $\alpha_s$, is not very
different at the $J/\psi$ and $\psi(2S)$ masses, it is expected that 
for any state $h$ the
 $J/\psi$ and $\psi(2S)$ branching ratios are related by

\begin{equation}
Q_h=\frac{{\cal B}(\psi(2S)\to h)}{{\cal B}(J/\psi\to h)}
\approx
\frac{{\cal B}(\psi(2S)\to \ell^+\ell^-)}{{\cal B}(J/\psi\to\ell^+\ell^-)}= (12.7 \pm 0.5)\%,
\end{equation}
where ${\cal B}$ denotes a branching fraction, $h$ is a particular hadronic final
state, and the leptonic branching fractions are taken from the PDG~\cite{PDG}.
This relation  is sometimes called
\lq \lq the 12\% rule''. Modest deviations from the rule are expected~\cite{GULI}. 
Although the rule works well for some specific decay modes of the 
$\psi(2S)$, it fails spectacularly for $\psi(2S)$ decays to final states 
consisting of one vector and one pseudoscalar meson (VP), such as $\rho\pi$. 

Values of $Q_h$ have been measured for a wide variety of final
states~\cite{PDG}. A recent
review~\cite{GULI} of relevant theory and experiment concludes that current
theoretical explanations are unsatisfactory.
Clearly more experimental results are desirable. This paper presents 
measurements of the following new decay modes of the $\psi(2S)$: 
$\pi^+\pi^-\pi^0$, $\rho\pi$, $\omega\pi$, $\rho\eta$, $K^{*0}(892)\bar{K^0}$, 
$\eta 3\pi$, $\eta^\prime 3\pi$, $2(K^+ K^-)$, $p \bar p K^+ K^-$,
$\Lambda \bar \Lambda \pi^+ \pi^-$, $\Lambda \bar p K^+$,
$\Lambda \bar p  K^+ \pi^+ \pi^-$,
and more precise measurements of these previously measured modes: 
$\phi\pi$, $\omega\eta$, $\phi\eta$, $K^{*+}(892)K^-$, $b_1(1235)\pi$,
$2(\pi^+ \pi^-)$, $2(\pi^+ \pi^-) \pi^0$, $\omega \pi^+ \pi^-$,
$K^+ K^- \pi^+ \pi^-$, $\phi \pi^+ \pi^-$, $\omega K^+ K^-$, $\phi K^+ K^-$,
$p \bar p \pi^+ \pi^-$, $\omega p \bar p$, and $\Lambda \bar \Lambda$.
We also obtain an improved upper limit for $\phi p \bar p$.
Where applicable, the inclusion of charge conjugate modes is implied.
We select intermediate final states in the following decay modes:
 $\rho\to\pi\pi$, $\pi^0\to\gamma\gamma$,
$\omega\to\pi^+\pi^-\pi^0$, $\phi\to K^+K^-$,
$\eta\to\gamma\gamma$ and $\pi^+\pi^-\pi^0$, $K^{*0}\to K^-\pi^+$,
$K^{*+}\to K_S^0\pi^+$ and $K^+\pi^0$, $K_S^0\to\pi^+\pi^-$ and
$\Lambda\to p \pi^-$.

\section{Analysis and Results}

The data sample used in this analysis is obtained at the $\psi(2S)$ and the
nearby continuum in $e^+e^-$ collisions produced by the Cornell Electron
Storage Ring (CESR) and acquired with the CLEO detector.
The datasets consist of $\cal{L}$=5.46~pb$^{-1}$ on the
peak of the $\psi(2S)$ (2.57~pb$^{-1}$ for CLEO~III~\cite{cleoiiidetector}, 
2.89~pb$^{-1}$ for CLEO-c~\cite{YELLOWBOOK}) and 20.5~pb$^{-1}$ at 
$\sqrt{s}$=3.67~GeV (all CLEO-c). The nominal
scale factor used to normalize continuum yields to $\psi(2S)$ data is
$f_{\rm nom}=0.2645\pm0.004$, and is determined from the integrated
luminosities of the data sets corrected for the $1/s$ dependence of the
cross section, where the error is from the relative luminosity uncertainty. 
The actual $f$ used for each mode also corrects for the small differences in
efficiency between the $\psi(2S)$ and continuum data samples.

Standard requirements are used to select charged particles reconstructed 
in the tracking system and photon candidates in the CsI calorimeter. 
Charged particles are identified using a likelihood function that 
incorporates dE/dx from the drift chamber and information from the RICH.
The invariant mass of the decay products from intermediate resonant states
such as $\pi^0$, $\eta$, $\rho$, $\omega$, $\phi$, $K^*$, $K_S$ and $\Lambda$ 
must lie within limits determined from Monte Carlo (MC) studies.

Energy and momentum conservation requirements are imposed on the
reconstructed final state hadrons, which have momentum $p_i$ and combined
measured energy $E_{\rm vis}$. We require the measured scaled energy
$E_{\rm vis}$/$E_{\rm cm}$ 
be consistent with unity within experimental resolution, which varies by final state. 
We require $|\Sigma {\bf p_i}|/E_{\rm cm}< 0.02$. Together these requirements
suppress backgrounds with missing energy or incorrect mass assignments. 
The experimental resolutions are
smaller than 1\% in scaled energy and 2\% in scaled momentum.

For the modes with two or more charged pions, two $\pi^0{\rm's}$ 
or an $\eta$, we reject $\psi(2S) \rightarrow J/ \psi X$ 
($X= \pi^+ \pi^-$, $\pi^0 \pi^0$, or $\eta$) events in which the
mass of any of the following falls within the range
$3.05<m<3.15$~GeV: the two highest momentum oppositely charged tracks, the
recoil mass against the two lowest momentum oppositely charged tracks, or
the mass recoiling against the 2$\pi^0{\rm's}$ or $\eta$.

For every  final state, a signal selection range in 
$E_{\rm vis}$/$E_{\rm cm}$ is determined by MC simulation,
and a sideband selection range is defined to measure background.  
Final states with intermediate particles must satisfy a scaled energy signal
selection range requirement identical to the corresponding mode without the
intermediate particle; and the event yield is determined from signal and
sideband selection ranges of the intermediate particle mass. 

The number of events attributable to each $\psi(2S)$ decay mode is
obtained by subtracting the number of events in the sideband region and
the scaled continuum production from the number of events in the signal 
region. Interference between $\psi(2S)$ decay and continuum production 
of the same final state is neglected.
The efficiency for each final state is the 
average obtained from MC simulations for both detector configurations;
the two values are typically within a few percent (relative) to each other.
No initial state radiation is included in the MC, but final state 
radiation is accounted for.

We correct the number of signal for the efficiency and normalize to the  
number of $\psi(2S)\to\pi^+\pi^-J/\psi$, $J/\psi\to\mu^+\mu^-$ decays
in the data, which has been determined previously to be 
$(6.75 \pm 0.12)\times 10^4$~\cite{BKHMK}.
The resulting relative  branching ratios are obtained and then converted
to the absolute branching ratios using 
${\cal B} (\psi(2S)\to\pi^+\pi^-J/\psi) =  0.323 \pm 0.013$ 
~\cite{PDG} and 
${\cal B} (J/\psi\to\mu^+\mu^-)=  (5.88 \pm 0.10)\% $~\cite{PDG}. 
$Q_h$ values are determined using the absolute $\psi(2S)$ branching ratios
determined in this analysis and $J/\psi$ branching ratios from \cite{PDG}.


The branching ratios and $Q_h$ values for 2-body modes are given in 
Table~\ref{tab:2body}, where BES numbers are from \cite{BESnew};
preliminary results for multi-body modes ~\cite{psipmulti} are listed 
in Table~\ref{tab:multi}.

\begin{table*}[h]
\begin{center}
\tbl{$\psi(2S)$ branding ratios to two-body final states and 
     a comparison to BES and the PDG.}
{\begin{tabular}{|l|r|r|c|} \hline
Mode              & CLEO $(10^{-5})$           &PDG/BES$(10^{-5})$& $Q_h$ (\%) \\
\hline
$\pi^+\pi^-\pi^0$ & $17.7^{+1.5}_{-1.3}\pm2.7$ &   $18.1\pm1.8\pm1.9$ & $0.8\pm0.1$ \\
\hline
$\rho^0\pi^0$     & $0.9^{+0.4}_{-0.4}\pm0.1$  &                      & $0.2\pm0.1$ \\
$\rho^+\pi^-$     & $1.0^{+0.6}_{-0.5}\pm0.1$  &                      & $0.1\pm0.1$ \\
\hline
$\rho\pi$         & $2.0^{+0.7}_{-0.6}\pm0.2$  &   $5.1\pm0.7\pm0.8$  & $0.2\pm0.1$ \\
$\omega\pi$       & $2.3^{+1.1}_{-0.9}\pm0.2$  &   $1.87^{+0.68}_{-0.62}\pm.28$ & $5.6\pm2.7$ \\
$\phi\pi$         & $<0.6$                     &   $<0.3$             & \\
$\rho\eta$        & $2.7^{+0.9}_{-0.8}\pm0.2$  &   $1.78^{+0.67}_{-0.62}\pm.17$ &$13.8\pm5.0$ \\
$\omega\eta$      & $<1.0$                     &   $<1.1$             & \\
$\phi\eta$        & $1.8^{+1.5}_{-1.0}\pm0.4$  &   $3.5\pm1.0\pm0.6$  & $2.8\pm2.3$ \\
$K^{*0}\bar{K}^0$ & $8.7^{+2.5}_{-2.1}\pm0.8$  &   $15.0\pm2.1\pm1.9$ & $2.1\pm0.6$ \\
$K^{*+}K^-$       & $1.0^{+0.9}_{-0.6}\pm0.2$  &   $2.9\pm1.3\pm0.4$  & $0.2\pm0.2$ \\
\hline
$b_1^0\pi^0$      & $20.5^{+4.4}_{-3.8}\pm2.9$ &                      & $8.9\pm3.0$ \\
$b_1^+\pi^-$      & $36.8\pm4.0\pm7.4$         & $32\pm8$             &$12.3\pm2.5$ \\
\hline
$b_1\pi$          & $56.6^{+5.5}_{-5.3}\pm10.8$&                      &$10.7\pm1.9$ \\
\hline
\end{tabular}}
\label{tab:2body}
\end{center}
\end{table*}

\begin{table*}[h]
\begin{center}
\tbl{$\psi(2S)$ branding ratios to multi-body final states, $h$, 
and the corresponding branching ratios at the $J/\psi$ from PDG, 
and $Q_h$ from Equation (1).}
{\begin{tabular}{|c|c|c||c|c|}  \hline
   mode             &\multicolumn{2}{c||}{${\cal B}(\psi(2S)\to h) \ (10^{-4})$} & $Br(J/\psi)$ $(10^{-4})$ & $Q_h$ (\%)\\ 
   $h$ &this work & PDG  & PDG & this work \\ \hline
$2(\pi^+ \pi^-)$              &  2.0 $\pm$ 0.3 & $4.5 \pm 1.0$ &  40    $\pm$ 10    &   5.0 $\pm$1.5\\
$2(\pi^+ \pi^-)\pi^0$         & 23.7 $\pm$ 3.3 & $ 30 \pm  8 $ & 337    $\pm$ 26    &   7.0 $\pm$1.1\\
$\omega \pi^+ \pi^-$          &  8.0 $\pm$ 1.2 & $4.8 \pm 0.9$ &  72    $\pm$ 10    &  11.1 $\pm$2.3\\
$\eta 3\pi$                   &  8.5 $\pm$ 1.0 & -             &          -         &          -      \\
$\eta^\prime 3\pi$            &  4.3 $\pm$ 2.0 & -             &          -         &          -      \\
$K^+K^-\pi^+\pi^-$            &  6.5 $\pm$ 0.9 & $ 16 \pm  4 $ &  72    $\pm$ 23    &   9.0 $\pm$3.1\\
$\phi \pi^+ \pi^-$            &  0.9 $\pm$ 0.2 & $1.50\pm0.28$ &   8.0  $\pm$  1.2  &  11.4 $\pm$3.5\\
$\omega K^+ K^-$              &  1.9 $\pm$ 0.4 & $1.5 \pm 0.4$ &  19    $\pm$  4    &   9.9 $\pm$2.9\\
$2(K^+K^-)$                   &  0.6 $\pm$ 0.1 & -             &          -         &          -      \\
$\phi K^+ K^-$                &  0.7 $\pm$ 0.2 & $0.60\pm0.22$ &  15.4  $\pm$  2.1  &   4.7 $\pm$1.6\\
$p \bar{p} \pi^+ \pi^-$       &  5.4 $\pm$ 0.7 & $8.0 \pm 2.0$ &  60    $\pm$  5    &   9.0 $\pm$1.4\\
$\omega p \bar{p}$            &  0.5 $\pm$ 0.2 & $0.80\pm0.32$ &  13    $\pm$  2.5  &   3.7 $\pm$1.9\\
$p \bar{p} K^+ K^-$           &  0.2 $\pm$ 0.1 & -             &          -         &          -      \\
$\phi p \bar{p}$              & $<0.18$(90\%CL)& $<0.26$       &   0.45 $\pm$  0.15 &          -      \\
$\Lambda \bar\Lambda$         &  3.0 $\pm$ 0.5 & $1.81\pm0.34$ &  13    $\pm$  1.2  &  23.4 $\pm$4.6\\
$\Lambda\bar\Lambda\pi^+\pi^-$&  2.7 $\pm$ 0.8 & -             &          -         &          -      \\
$\Lambda \bar{p} K^+$         &  0.7 $\pm$ 0.2 & -             &   8.9  $\pm$  1.6  &   7.9 $\pm$2.8\\
$\Lambda\bar{p}K^+\pi^+\pi^-$ &  1.2 $\pm$ 0.4 & -             &          -         &          -      \\ \hline
\end{tabular}}
\label{tab:multi}
\end{center}
\end{table*}

\vspace{-0.5cm}
\section{Summary}
We have presented first evidence for $\psi(2S)$
decays to $\pi^+\pi^-\pi^0$, $\rho\pi$, $\omega\pi$, $\rho\eta$, and 
$K^{*0}\bar{K^0}$. Measurements for several other VP channels are also given.
The results suggest that, for VP final states, $\psi(2S)$ decays
through three gluons are severely suppressed with respect to the 12\%
rule and the corresponding electromagnetic processes are not.
The decay $\psi(2S)\to \pi^+\pi^-\pi^0$ exhibits a $\rho\to\pi\pi$ signal
but has a much larger component at higher $\pi\pi$ mass.
We also present preliminary branching fractions 
for seven new multi-body decay modes of the $\psi(2S)$, namely
$\eta 3\pi$, $\eta^\prime 3\pi$, $2(K^+ K^-)$, $p \bar p K^+ K^-$,
$\Lambda \bar \Lambda \pi^+ \pi^-$, $\Lambda \bar p K^+$,
$\Lambda \bar p  K^+ \pi^+ \pi^-$, and
more precise measurements of nine previously measured modes, which are
$2(\pi^+ \pi^-)$, $2(\pi^+ \pi^-) \pi^0$, $\omega \pi^+ \pi^-$,
$K^+ K^- \pi^+ \pi^-$, $\phi \pi^+ \pi^-$, $\omega K^+ K^-$, $\phi K^+ K^-$,
$p \bar p \pi^+ \pi^-$, $\omega p \bar p$, and $\Lambda \bar \Lambda$. 
We also obtain an improved upper limit for $\phi p \bar p$.
For multi-body decay modes $Q_h$ values are in general lower than,
but, in most cases, consistent with the 12\% rule.

\section*{Acknowledgments}

We gratefully acknowledge the effort of the CESR staff 
in providing us with
excellent luminosity and running conditions.
This work was supported by 
the National Science Foundation,
the U.S. Department of Energy,
the Research Corporation,
and the Texas Advanced Research Program.


\end{document}